\documentstyle[aps,epsfig,prb,preprint]{revtex}
\begin{document}
\draft

\title{Strong-coupling Expansions at Finite Temperatures: Application
to Quantum Disordered and Quantum Critical Phases}
\author{Norbert Elstner}
\address{Physikalisches Institut, Universit\"at Bonn, 
         Nu\ss allee 12, D-53115 Bonn, Germany }
\author{Rajiv R. P. Singh}
\address{Department of Physics, University of California, 
Davis, California 95616}
\maketitle

\begin{abstract}
By combining conventional finite-temperature many-body perturbation theory with
cluster expansions, we develop a systematic method to carry out
high order arbitrary temperature perturbative calculations on the computer.
The method is well suited to studying the thermodynamic properties
of quantum disordered and quantum critical phases at finite temperatures.
As an application, we calculate the magnetic susceptibility, internal energy and
specific heat of the bilayer Heisenberg model. It is shown that for
a wide range of coupling constants these expansions show excellent convergence
at all temperatures. Comparing the direct series ( without extrapolations)
for the bulk susceptibility to Quantum Monte Carlo simulations we find 
an almost perfect agreement between the two methods even at the quantum 
critical coupling separating the dimerized and antiferromagnetic phases. 
The convergence fails only at very low temperatures, which are also
difficult to reach by Quantum Monte Carlo simulations.
\end{abstract}

\section{Introduction} 

Quantum critical phenomena and quantum disorder
in strongly correlated many-body systems have been a subject of
much interest lately. They have been invoked to account
for many properties of high temperature superconductors \cite{hightc},
for non fermi-liquid behavior in heavy fermion materials \cite{heavyf},
and for a variety of other interesting systems ranging from
superconductor-insulator transitions to the quantum hall effect \cite{sondhi}.
In many problems it is not possible to fine tune the system
parameters to be right at the quantum critical point. Nevertheless,
the quantum critical point controls their behavior 
above some very low temperature scale \cite{chn,CSY94,review97}.
The aim of our study is to develop a method
that will allow us to calculate the temperature dependence
of thermodynamic quantities with high
accuracy for many model Hamiltonians, when the system parameters are in
the quantum disordered phase or
near a zero temperature quantum critical point.

Quantum spin-systems provide a particularly rich variety
of models and real materials where quantum critical phenomena
can be studied. 
There exist many experimental realizations of spin-ladders
and alternating chains \cite{dagotto-rice}. 
Bilayer Heisenberg model is considered relevant
to high temperature superconductors \cite{ybco}.
Recently, the material
$CaV_4O_9$ has attracted considerable interest \cite{cavo}.
The spins in these materials are
arranged in a novel geometrical arrangement, which
allows for many possible types of spin-disordered ground states
and quantum critical points.

Finite temperature properties of such systems have been studied before by
finite-size studies, including exact diagonalization \cite{prelovsek} and
Quantum Monte Carlo simulations \cite{sandvik94,sandvik95}, 
and by high temperature expansions \cite{weihong}.
The latter method performs poorly at low temperatures for many
parameter regimes of interest. Exact diagonalization, in higher
than one dimension, is generally limited to fairly small systems
and extrapolation to thermodynamic limit is unreliable.
The Quantum Monte Carlo method
is perhaps the most accurate to date, but it suffers from the
sign problem when frustration is present. 

Here, we combine many-body perturbation theory with cluster
expansions to develop high-order strong-coupling expansions for
thermodynamic quantities, such as the bulk susceptibility
and the specific heat, at arbitrary temperatures.
The resulting series are
perturbation expansions in the weaker couplings, whose coefficients
depend on temperature. For the bilayer Heisenberg model,
these series show excellent convergence at
all temperatures for a range of parameters ( without using
series extrapolation methods such as Pade approximants). Even
when the system parameters are tuned to the quantum critical
point, the series converges very well down to fairly low temperatures,
comparable to the lowest temperatures accessible in the Monte
Carlo simulations.

The method should prove useful for frustrated as well as quenched
random spin syatems. It can also be applied to electronic models
such as t-J and Hubbard or Kondo lattice models. Being a finite temperature 
perturbation expansion, it is not limited to systems with
non-degenerate ground states.

The plan of the paper is as follows: In section II we discuss the
basic perturbation theory. In section III we introduce some techniques
that are needed for an effective computer implementation of the
method. In section IV the series expansions for the bilayer
Heisenberg model is presented. In section V we compare our series
with Quantum Monte Carlo simulations. Finally, in section VI
we present our conclusions and suggest future directions.

\section{Perturbation Theory} 

We are interested in systems which are described by a Hamiltonian
\begin{equation}
   {\cal H}= H_0+\lambda H_1 \;\; .
\end{equation}
Here, $H_0$ consists of all couplings within an elementary cluster, whereas,
$H_1$ describes interactions between different such units.
The basic idea of cluster expansions is to exploit the fact that in 
any finite order of perturbation theory only a finite number of these 
elementary units can be coupled together and thus expansions for a 
thermodynamic system can be obtained exactly by carrying out perturbation 
theory just for these few finite-clusters or graphs ( In this paper we will
refer to them as graphs to distinguish them from the elementary
clusters which define the Hamiltonian $H_0$). A thermodynamic quantity,
such as magnetic susceptibility per elementary cluster, $\chi$, 
for the lattice ${\cal L}$ can  be written as
\begin{equation}
   \chi ({\cal L})=\chi_0+ \sum_g L(g)\times W(g) \;\; .
\end{equation}
Here $\chi_0$ is the susceptibility of the elmentary cluster for $\lambda = 0$ 
and the sum runs over all connected graphs of the lattice. The quantity
$L(g)$, called the lattice constant of the graph, is defined as the number
of ways per elementary cluster that the graph $g$ can be embedded
in the lattice. The quantity $W(g)$ is the weight of the graph $g$
defined by the recursion relation,
\begin{equation}
   W(g)= \chi(g)-N_g\chi_0-\sum_{g^\prime} W(g^\prime) \;\; ,
\end{equation}
where, $N_g$ is the number of sites in the graph and the
sum over $g^\prime$ runs over all proper subgraphs of the graph $g$.
For a graph $g$ with $B_g$ bonds it can be shown \cite{GSH90} that 
$W(g) = O(\lambda^{B_g})$. 

So far the formalism is identical to a standard high temperature
expansion. 
The novelty arises in the calculation of $\chi(g)$
for a finite graph. In order to do that, we only need
to consider the spins in that graph and the interactions
$H_0$ and $H_1$ between these spins.  To calculate expansions at
arbitrary temperature, we exploit the following relation
\begin{equation}
   e^{-\beta(H_0+\lambda H_1)} = e^{-\beta H_0} \sum_n (-\lambda)^n I_n\;\; ,
\end{equation}
where $I_n$ are n-fold integrals given by,
\begin{equation}
   I_n=\int_0^\beta dt_1\int_0^{t_1} dt_2 \ldots H_1(t_1)H_1(t_2)\ldots 
\end{equation}
Here, the operators, $H_1(t)$ have the standard time dependence of
the interaction representation
\begin{equation}
   H_1(t)=e^{tH_0}H_1e^{-tH_0} \;\; .
\end{equation}
Thus, the partition function can be reduced to the expression,
\begin{equation}
   Z=Z_0+\sum_{n=1} (-\lambda)^n Z_n \;\; , 
\end{equation}
where $Z_n$ are given by
\begin{equation}
   \label{Z_n}
   Z_n \! = \! \int_0^\beta \! dt_1\int_0^{t_1} \! dt_2 \ldots 
   Tr [e^{-\beta H_0} H_1(t_1)H_1(t_2)\ldots] \;\; .
\end{equation}
It is evident that in order to evaluate these expressions,
we need to work in a basis in which $H_0$ is diagonal.
This basis is simply a direct product of the eigen-basis for
elementary clusters. Thus calculating thermodynamic quantities
is straightforward if the matrix elements of $H_0$ and
$H_1$ are known in this basis. In the next section we
discuss an efficient method for calculating these traces and
integrals.

\section{ Evaluating the Multiple Integrals}

The basic energy scale in this problem is set by the level spacing $\Delta_0$ 
in the spectrum of the unperturbed part $H_0$. Measuring temperature 
in units of this quantity one finds that calculating $Z_n$ requires 
repeated integrations over functions of type:
\begin{equation}
   I(k,l;x_{\nu}) =  x_{\nu}^k \; e^{lx_{\nu}}
\end{equation}
Where $k$ and $l$ are integers and $k \ge 0$. 
It is then easy to see that the $I(k,l;x)$ form a closed set, because

\begin{eqnarray}
   l \ne 0: \hskip 0.6in & & \nonumber \\
   \int_0^{x_{\nu-1}} \!\!\!\! dx_{\nu} \; x_{\nu}^k \; e^{lx_{\nu}}
   &=& k! \; \left( \frac{-1}{l} \right)^{k+1} \\
   &+& \sum_{i=0}^k \; (-1)^i \; \frac{1}{l^{i+1}} \; 
   \frac{k!}{(k-i)!} \; x_{\nu-1}^{k-i} \; e^{l x_{\nu-1}} \nonumber \\ 
   l = 0: \hskip 0.6in & & \nonumber \\
   \int_0^{x_{\nu-1}} dx_{\nu} \; x_{\nu}^k 
   &=& \frac{1}{k+1} \; x_{\nu-1}^{k+1} 
\end{eqnarray}

These equations allow for an iterative evaluation of the multiple
integrals entering Eqn.(\ref{Z_n}). One finds that the coefficients
$Z_n$ are finite polynomials in the two variables $x = \Delta_0/k_{\rm B}T$ 
and $y = \exp\left( -\Delta_0/k_{\rm B}T \right) $. 

In the following we will apply this series expansion method to systems 
where the basic cluster consists of a pair of $s=1/2$ spins coupled by a 
Heisenberg exchange $J_{\perp}$. Thus, $\Delta_0$ is the singlet-triplet 
spacing in the spectrum of $H_0$. In this case it turns out to be more 
convenient to use the variables $(x,Z_0)$ instead of the variables $(x,y)$, 
where $Z_0 = 1 + 3y$, is the zeroth order partition function.

\section{ Model and Tables}

Here, we apply the method to the spin-half bilayer Heisenberg model,
defined by the Hamiltonian 
\begin{equation}
   \label{bilayer}
   {\cal H} = J_{\perp} \sum_{i} {\bf S}_{A,i} \cdot {\bf S}_{B,i} 
            + J_{\parallel} \sum_{\left<i,j\right>} 
                                       {\bf S}_{A,i} \cdot {\bf S}_{A,j}
                                     + {\bf S}_{B,i} \cdot {\bf S}_{B,j}
\end{equation}
Here, the index $i$ enumerates sites on a two dimensional 
square lattice and $\left<i,j\right>$ are pairs of nearest neighbour sites
on this lattice.
The expansion parameter is given by ratio of the inter dimer 
to the intra dimer coupling:
\begin{equation}
   \lambda = J_{\parallel}/J_{\perp} \;\; .
\end{equation}
The series for the susceptibility $\chi$ and the 
logarithm of the partition function $\ln Z$ per dimer
are presented in Tables 1 and 2. The series can also be obtained 
on the WWW. The acess adress is given at the end of this article.

\section{ Results and Comparisons}
In this section we show the convergence of the expansions by comparing
partial sums of different order and by comparing with the Quantum
Monte Carlo data. It is known from a number of studies at $T=0$ that
this model has a quantum critical point at $\lambda\approx 0.4$
\cite{sandvik94,hida}.
We show here results 
in the quantum disordered phase ( at $\lambda=0.3$ ) and near the
quantum critical point at $\lambda=0.4$.

\begin{figure}[ht]
  \protect\centerline{\epsfig{file=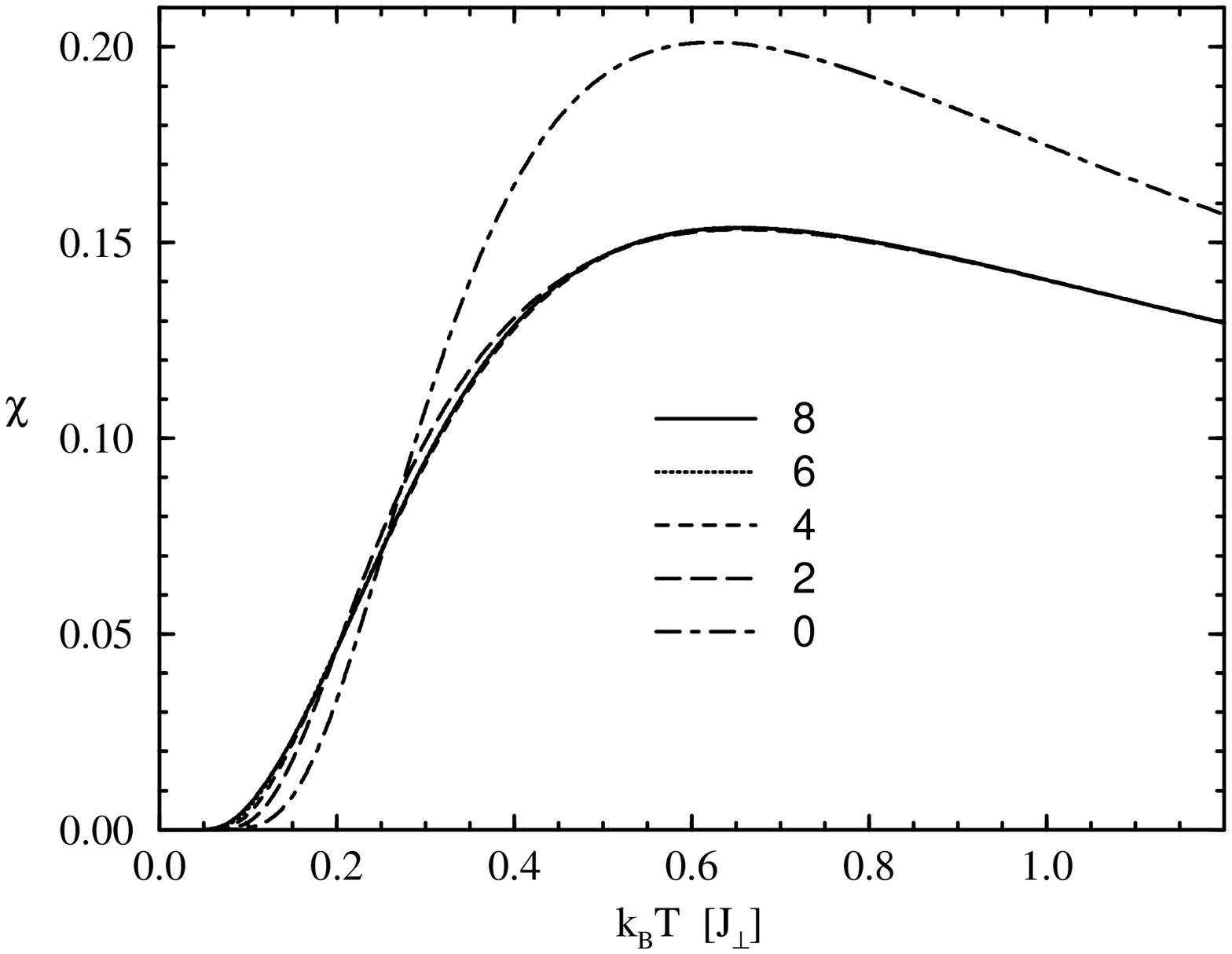, height=5in}}
  \protect\caption{ Susceptibility $\chi$ per spin vs. temperature 
  $k_{\rm B}T/J_{\perp}$ for $J_{\parallel}/J_{\perp} = 0.3$, 
  i.e. in the quantum disordered regime.  
  The lines are series of order 2, 4, 6 and 8 compared to the
  susceptibility of an isolated dimer (0). } 
\label{fig:chi_d}
\end{figure}

\begin{figure}[ht]
  \protect\centerline{\epsfig{file=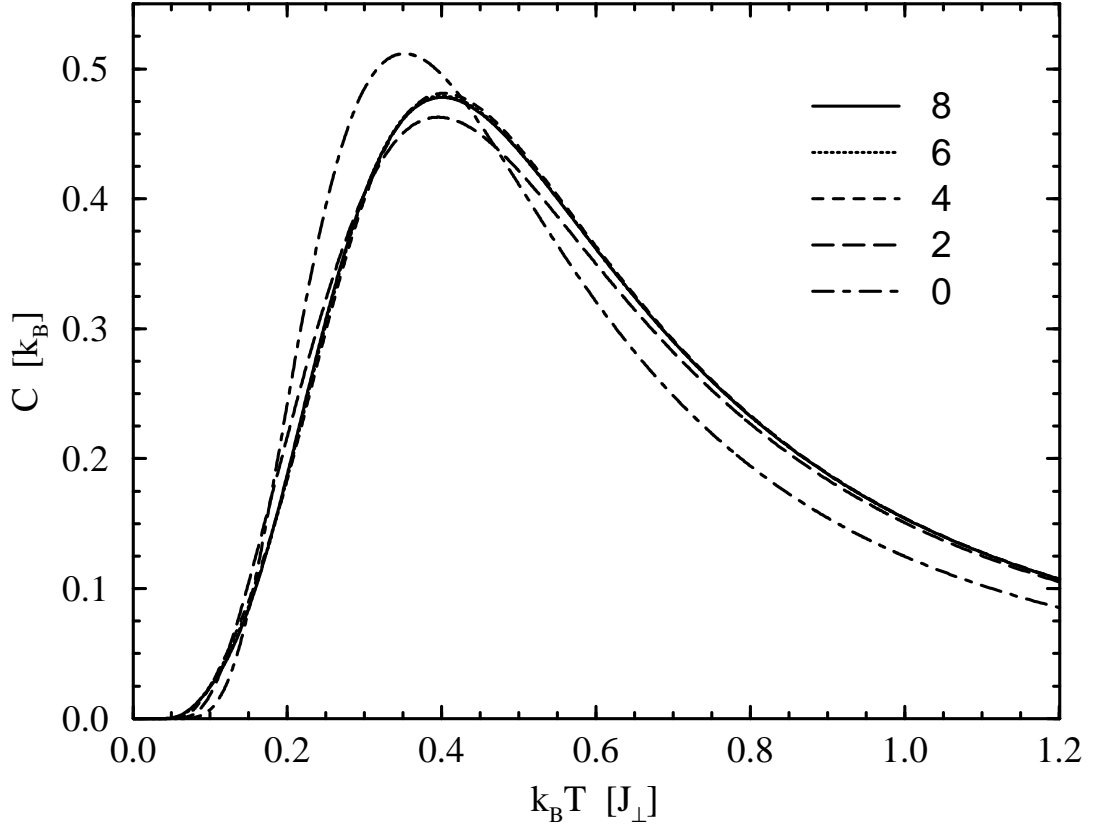, height=5in}}
  \protect\caption{ Specific heat $C$ per spin vs. temperature 
  $k_{\rm B}T/J_{\perp}$ for $J_{\parallel}/J_{\perp} = 0.3$, 
  i.e. in the quantum disordered regime.  
  The lines are series of order 2, 4, 6 and 8 compared to the
  specific heat of an isolated dimer (0). } 
\label{fig:C_d}
\end{figure}

In figures (\ref{fig:chi_d}) and (\ref{fig:C_d}) the susceptibility 
and specific heat are plotted as a function of temperature for 
$\lambda=0.3$. One can see that there is excellent convergence at 
all temperatures.

The quantum critical regime was investigated by Chubukov, 
Sachdev and Ye\cite{CSY94} by a large-N expansion of the 
quantum nonlinear $\sigma$-model. They obtained very detailed 
results for the low temperature behaviour. In particular the following 
predictions for the susceptibility and the specific heat {\sl per unit cell} 
were derived:
\begin{eqnarray}
 \chi &=& \frac{k_{\rm B}T}{(\hbar c)^2}\frac{\sqrt 5}{\pi} 
 \ln \left( \frac{\sqrt 5 + 1}{2} \right) 
 \left[ 1 - \frac{0.6189}{N} + O(N^{-2}) \right] \\
 C &=& \frac{3 \zeta (3)}{\pi} k_{\rm B} 
 \left( \frac{k_{\rm B}T}{\hbar c} \right)^2 
  N \left[ \frac{4}{5} - \frac{0.3344}{N} + O(N^{-2}) \right] 
\end{eqnarray}
Where $c$ is the spin wave velocity. From various zero temperature
calculations its value is known to be \cite{hida}:
\begin{equation}
C = 1.90 \; J_{\parallel}
\end{equation}
Evaluating these equations for $N=3$ gives for the susceptibility and 
specific heat {\sl per site}: 
\begin{eqnarray}
   \label{eqn:chi_qc}
 \chi &=& \frac{1}{2 \lambda^2} \; 0.272 \; k_{\rm B}T \\
   \label{eqn:C_qc}
 C &=& \frac{1}{2 \lambda^2} \; 2.371 \; k_{\rm B} \; (k_{\rm B}T)^2 
\end{eqnarray}
Here temperature is in units of the intra dimer coupling $J_{\perp}$. 

A very sensitive measure for quantum criticality is the Wilson ratio 
defined by
\begin{equation}
W = \frac{k_{\rm B}^2 T \; \chi(T)}{C(T)}.
\end{equation}
The numerical value of this dimensionless quantity follows immediately 
from Eqns.(\ref{eqn:chi_qc}) and (\ref{eqn:C_qc}): 
\begin{equation}
   \label{eqn:Wr_qc}
W = 0.115 \; \; \; \; . 
\end{equation}

\begin{figure}[ht]
  \protect\centerline{\epsfig{file=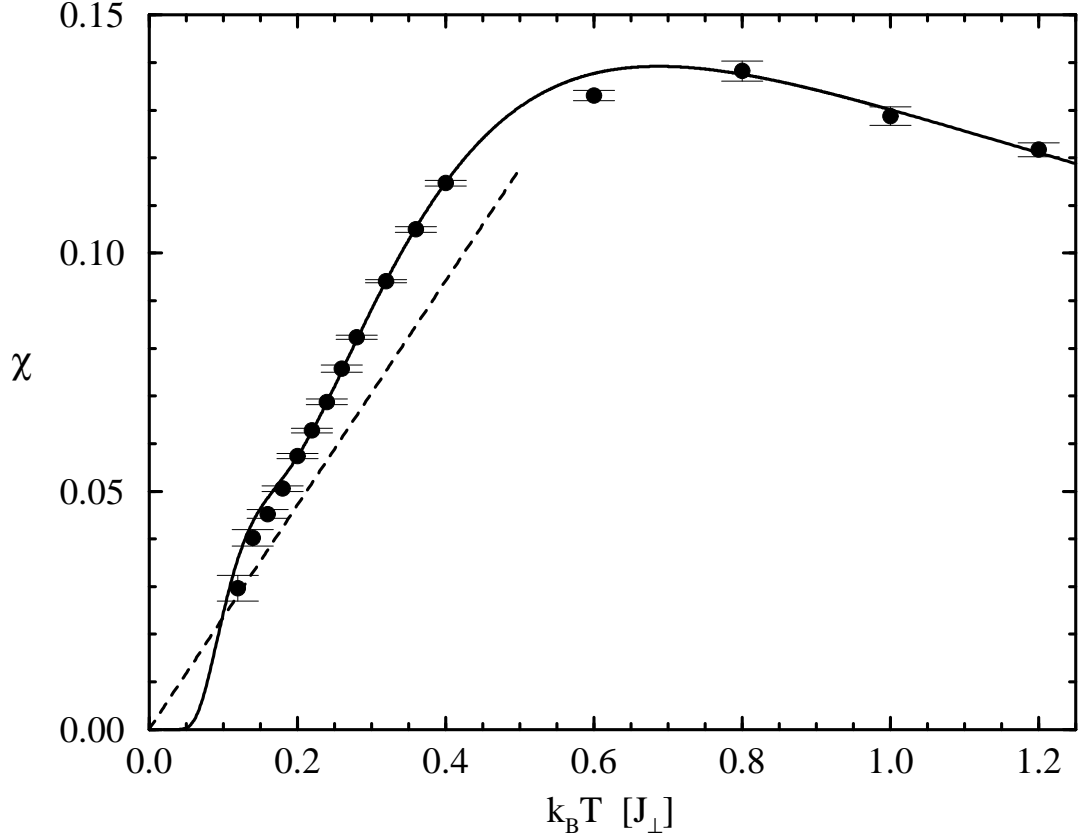, height=5in}}
  \protect\caption{ Susceptibility $\chi$ per spin vs. temperature 
  $k_{\rm B}T/J_{\perp}$ for $J_{\parallel}/J_{\perp} = 0.4$, 
  i.e. close to the quantum critical point. 
  The black circles show the QMC data of Sandvik and Scalapino. 
  The solid lines is the series of order 8. The dashed line is 
  the quantum critical (QC) prediction of Eqn. (17) . }
\label{fig:chi_qc}
\end{figure}

\begin{figure}[ht]
  \protect\centerline{\epsfig{file=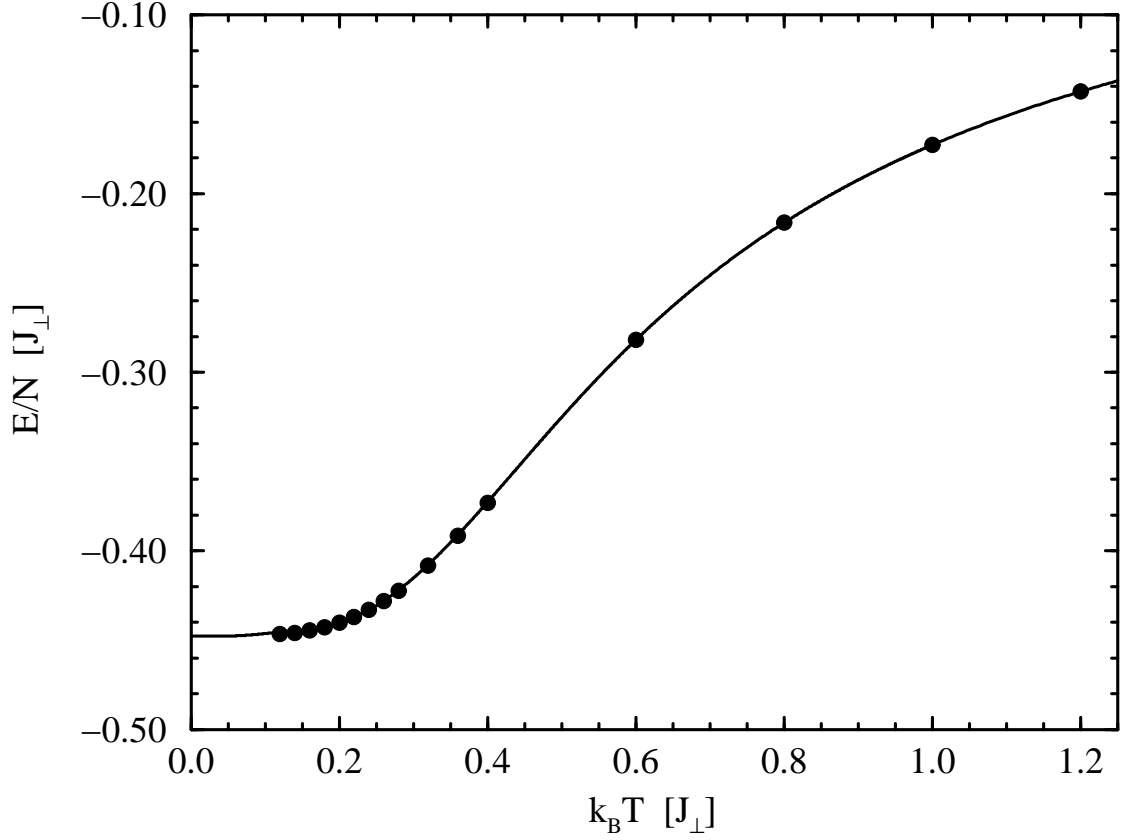, height=5in}}
  \protect\caption{ Internal Energy per spin $E/NJ_{\perp}$ vs. temperature 
  $k_{\rm B}T/J_{\perp}$ for $J_{\parallel}/J_{\perp} = 0.4$.
  The line is the series up to order 8. 
  The black dots show the QMC data of Sandvik and Scalapino. }
\label{fig:intE_qc}
\end{figure}

\begin{figure}[ht]
  \protect\centerline{\epsfig{file=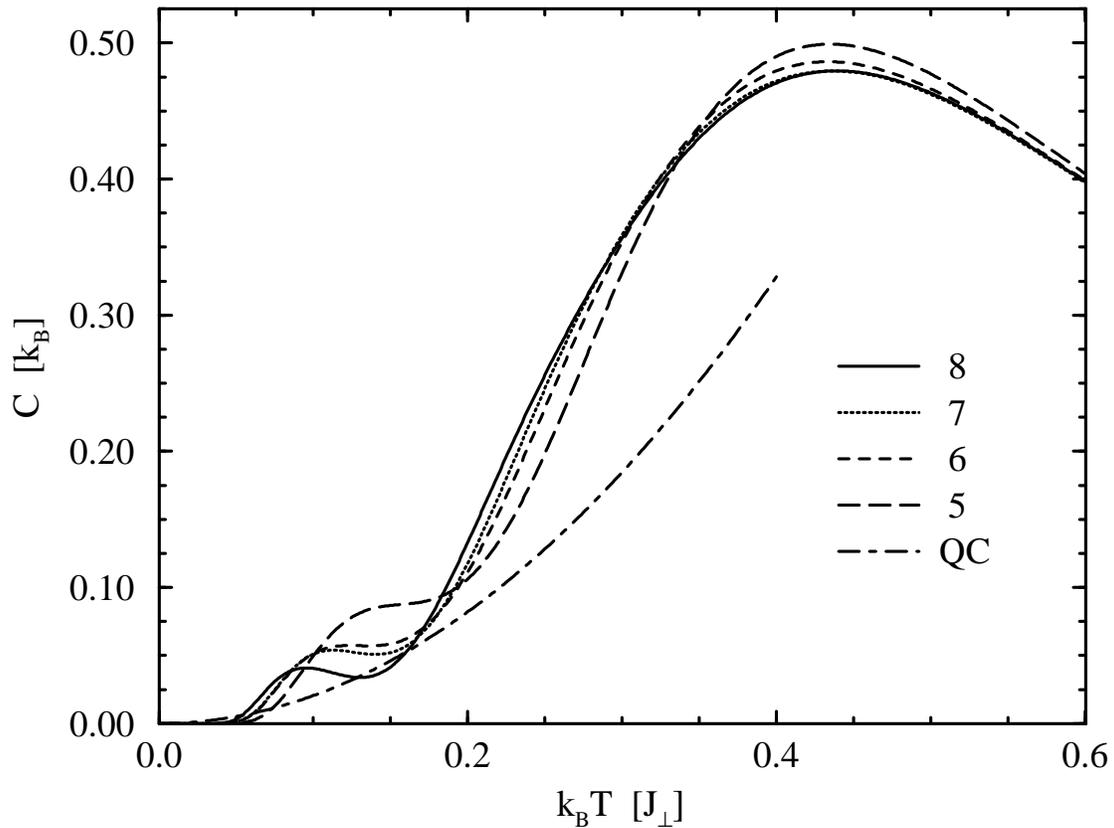, height=5in}}
  \protect\caption{ Specific heat per spin $C$ vs. temperature 
  $k_{\rm B}T/J_{\perp}$ for $J_{\parallel}/J_{\perp} = 0.4$.
  The lines are series of order 5, 6, 7 and 8 and 
  the quantum critical (QC) prediction of Eqn. (18) . }
\label{fig:Cv_qc}
\end{figure}

\begin{figure}[ht]
  \protect\centerline{\epsfig{file=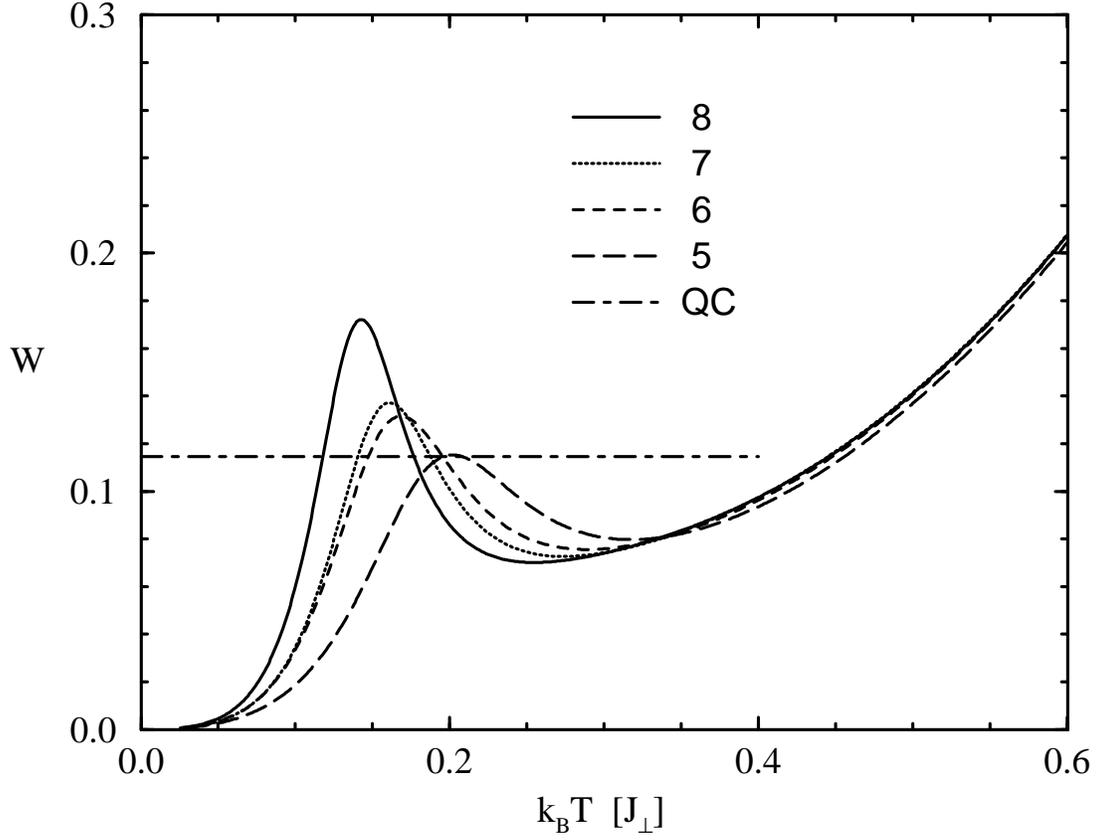, height=5in}}
  \protect\caption{ Wilson ratio $W$ vs. temperature 
  $k_{\rm B}T/J_{\perp}$ for $J_{\parallel}/J_{\perp} = 0.4$.
  $W$ obtained from series for $\chi$ and $C$ of order 5, 6, 7 and 8 
  and  compared to the quantum critical (QC) prediction 
  of Eqn. (20) . }
\label{fig:Wr_qc}
\end{figure}

In figures \ref{fig:chi_qc}, \ref{fig:intE_qc}, \ref{fig:Cv_qc},  
and  \ref{fig:Wr_qc},  the susceptibility, internal energy, specific heat, 
and Wilson ratio for the model are shown and compared with 
Monte Carlo data and with the asymptotic
quantum critical predictions. We see that the series converges
extremely well down to fairly low temperatures. However, the
convergence appears to break down just as the asymptotic
quantum critical behavior sets in. The finite series become
oscillatory at these very low temperatures. The activated behaviour 
at extremly low temperatures as seen in figures \ref{fig:chi_qc},
\ref{fig:Cv_qc} and \ref{fig:Wr_qc} is also an artifact of finite
series. In principle various extrapolation techniques, e.g. Pade approximants, 
can be used to enhance the convergence in this region. Here, we restrict 
the analysis to the temperature range where the series converge.

One interesting result to emerge from this study is that asymptotic
quantum critical scaling occurs in the model only at very low temperatures,
much lower than anticipated before. The uniform susceptibility
between $T=0.1J_\perp$ and $T=0.4J_\perp$ appears nearly linear but is not
strictly so.  There is a shoulder around $T=0.2J_\perp$, which is
present in both the series results and the Monte Carlo data.

On the other hand, many amplitude ratios follow the universal 
quantum critical predictions even above $T=0.1J_\perp$ \cite{sandvik95}.
Thus, the extent of the quantum critical regime in 2D systems 
appears to strongly depend on the quantity studied. Clearly, the role
of lattice corrections needs to be better understood \cite{EGSS95}.

\section{ Conclusions and Future Directions}

In this paper we presented a practical method to calculate high order
strong coupling expansions for quantum statistical models at
arbitrary temperatures. The method was applied to the bilayer
Heisenberg model, where it shows excellent convergence even
near the quantum critical point. In a companion paper,
we will discuss application of these methods to alternating
spin-chains and spin-ladders. The method is also applicable
to frustrated and quenched random spin models as well as to
Hubbard or Kondo models around the strong coupling limit. 
We are preparing to pursue these calculations in the future.

Series for the susceptibility $\chi$, internal energy $E$
and specific heat $C$ may be obtained on the WWW. In addition 
a Fortran program to read the data files and sum the series
is also available. 

The WWW access is via
http://brahms.physik.uni-bonn.de/\~\ norbert/series/series.html

Acknowledgements: 

We are greatful to A. Sandvik for providing the QMC data.
One of us (NE) acknowledges the hospitality of the University 
of California at Davis where part of this work was done.
This work is supported in part by the US National Science Foundation 
under Grant No. DMR-96-16574 (RRPS).

\begin{figure}[ht]
  \protect\centerline{\epsfig{file=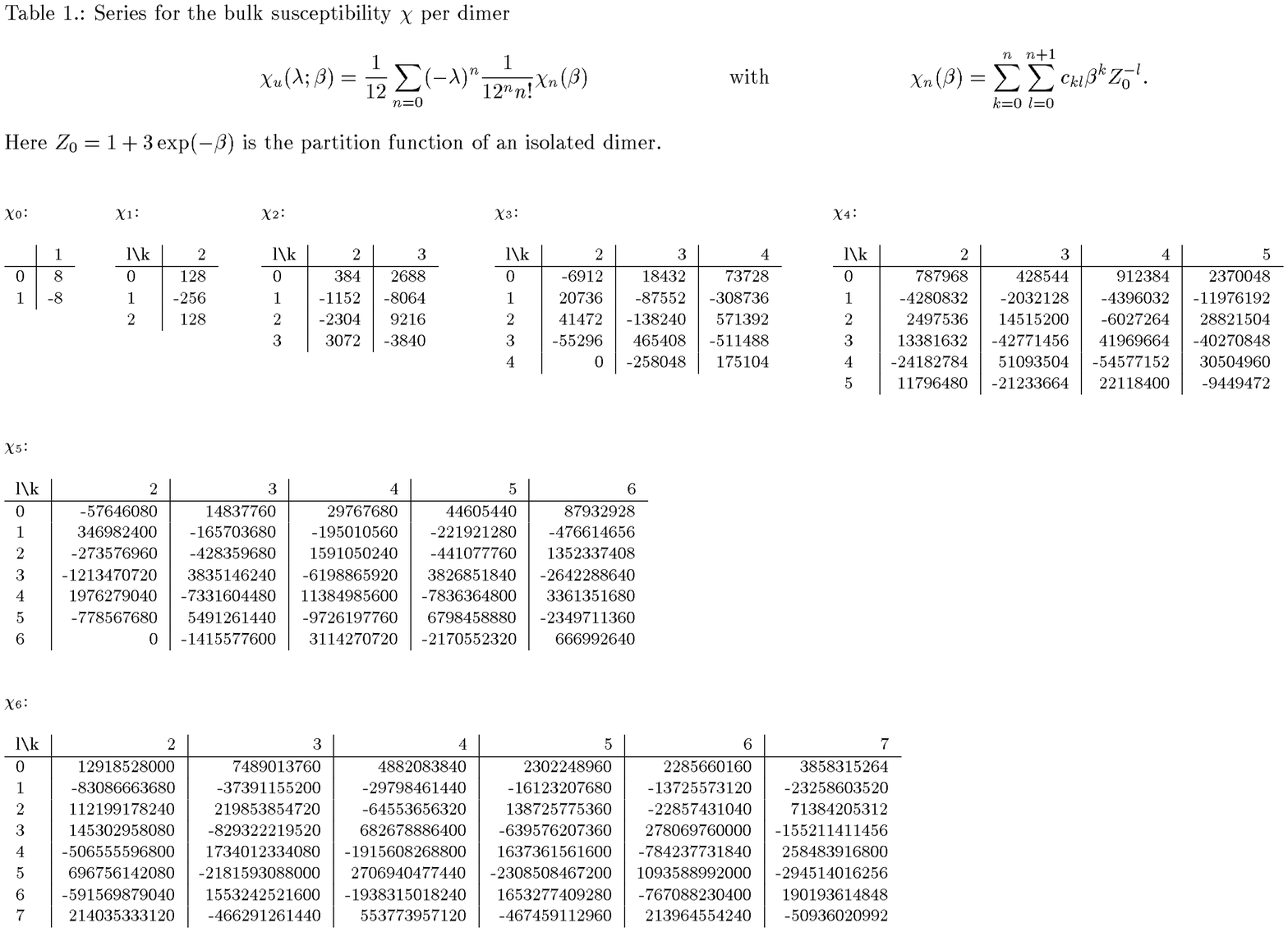, height=5.9in, angle=-90}}
\end{figure}

\begin{figure}[ht]
  \protect\centerline{\epsfig{file=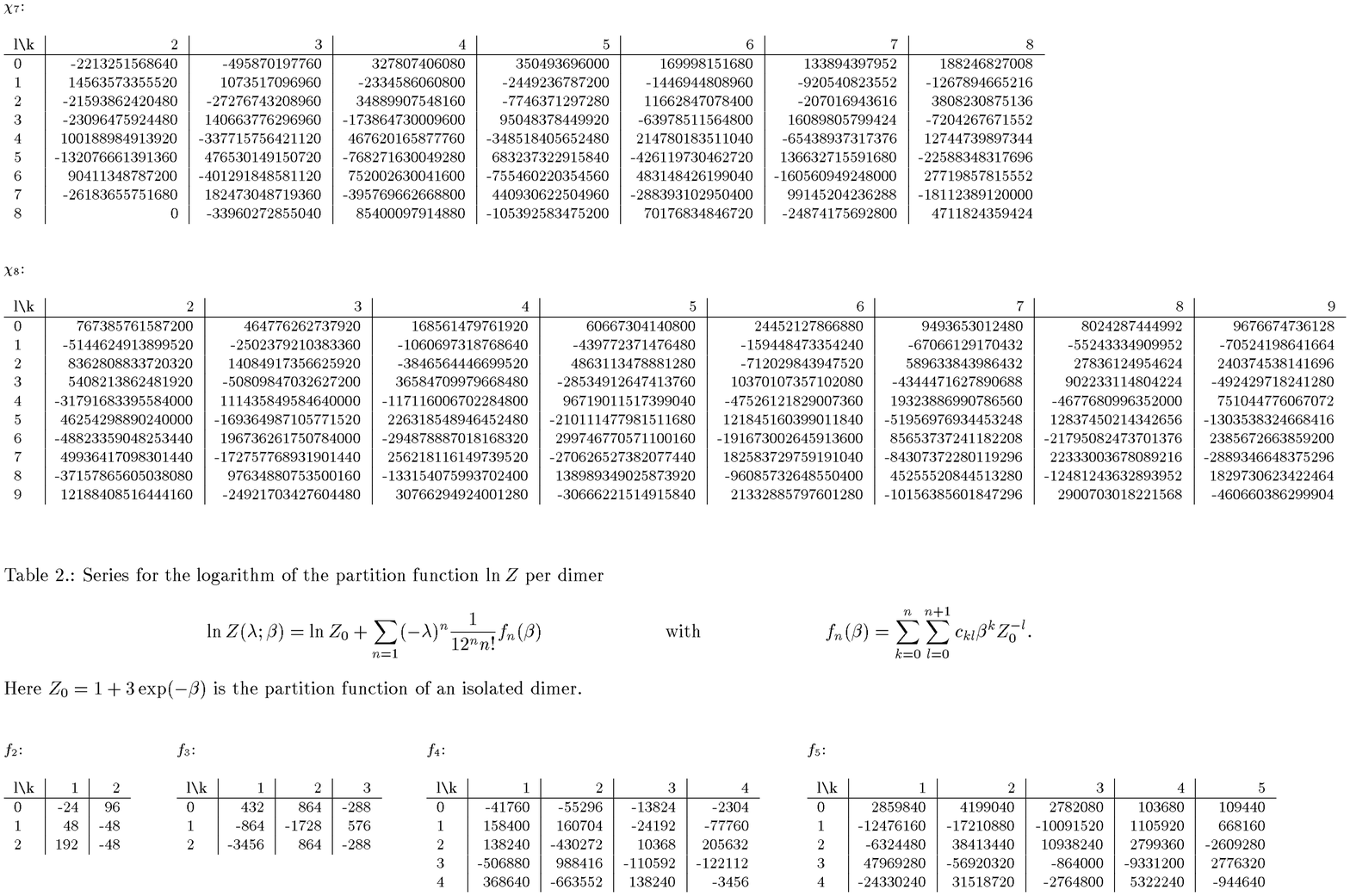, height=5.9in, angle=-90}}
\end{figure}

\begin{figure}[ht]
  \protect\centerline{\epsfig{file=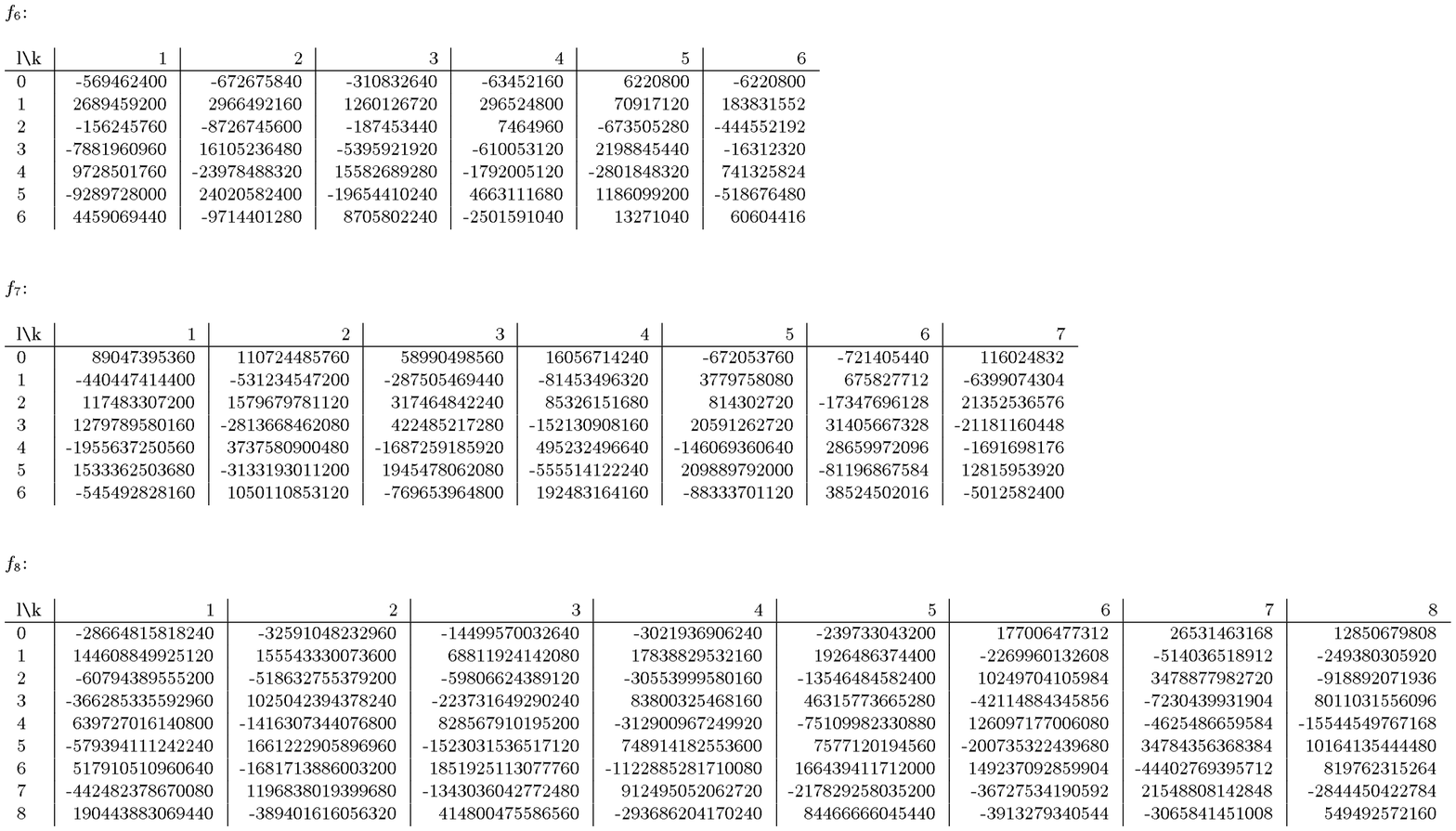, height=5.9in, angle=-90}}
\end{figure}


\begin{references}

\bibitem{hightc} A. Sokol and D. Pines, Phys. Rev. Lett. {\bf 71},
2813 (1993); A. V. Chubukov, S. Sachdev and A. Sokol, Phys. Rev. B{\bf 49},
9052 (1994). 

\bibitem{heavyf} M. C. Aronson {\it et al.}, 
Phys. Rev. Lett. {\bf 75}, 725 (1995).

\bibitem{sondhi} S. L. Sondhi, S. M. Girvin, J. P. Carini and D. Sahar, 
Rev. Mod. Phys. {\bf 69}, 315 (1997).

\bibitem{chn} S. Chakravarty, B. I. Halperin and D. R. Nelson,
Phys. Rev. B {\bf 39}, 2344 (1989).

\bibitem{CSY94} A. V. Chubukov, S. Sachdev and J. W. Ye, Phys. Rev.
B {\bf 49}, 11919 (1994); S. Sachdev and J. W. Ye, Phys. Rev. Lett.
{\bf 69}, 2411 (1992). 

\bibitem{review97} N. Elstner, Int. J. Mod. Phys. B {\bf 11}, 1753 (1997).

\bibitem{dagotto-rice} E. Dagotto and T. M. Rice, Science {\bf 271},
618 (1996).

\bibitem{ybco} A. J. Millis and H. Monien, Phys. Rev. B{\bf 50}, 
16606 (1994); H. Monien and T. M. Rice, Physica C, 1705 (1994).

\bibitem{cavo}
K. Udea {\it et al.}
Phys. Rev. Lett. {\bf 76}, 1932 (1996);
M. Troyer {\it et al.}
Phys. Rev. Lett. {\bf 76}, 3822 (1996).
O. A. Starykh {\it et al.}, Phys. Rev. Lett. {\bf 77},
2558 (1996).
M. P. Gelfand {\it et al.}, 
Phys. Rev. Lett. {\bf 77}, 2794 (1996).
W. E. Pickett,  Phys. Rev. Lett. {\bf 79}, 1746 (1997).

\bibitem{prelovsek} J. Jaklic and P. Prelovsek, Phys. Rev. Lett.
{\bf 77}, 892 (1996).

\bibitem{sandvik94}
A. W. Sandvik and D. J. Scalapino, 
Phys. Rev. Lett. {\bf 72}, 2777 (1994)

\bibitem{sandvik95} A. W. Sandvik, A. V. Chubukov and S. Sachdev,
Phys. Rev. B{\bf 51}, 16483 (1995).

\bibitem{weihong} J. Oitmaa {\it et al.}, {\bf 54}, 1009 (1996);
W. H. Zheng {\it et al.}, Phys. Rev. B {\bf 55}, 11377 (1997).

\bibitem{GSH90} M. P. Gelfand, R. R. P. Singh and D. A. Huse, 
J. Stat. Phys. {\bf 59}, 1093 (1990).

\bibitem{hida} K. Hida, J. Phys. Soc. Jpn {\bf 61}, 1013 (1992);
M. P. Gelfand, Phys. Rev. B {\bf 53}, 11309 (1996); W. H. Zheng
Phys. Rev. B{\bf 55}, 12267 (1997).

\bibitem{EGSS95} N. Elstner, R. L. Glenister, R. R. P. Singh and 
A. Sokol, Phys. Rev. B{\bf 51}, 8984 (1995).

\end{references}
\end{document}